\pdfoutput=1
\documentclass{optica-article}
\usepackage[utf8]{inputenc}

\journal{ome} % for journals or Optica Open

\articletype{Research Article}

\usepackage{amsmath}
\usepackage{graphicx}
\usepackage{xcolor}
\usepackage{braket}
\begin{document}

\title{Spin-polarized lasing in a photonic lattice}

\author{A. Herrero Otermin,\authormark{1,*} N. Carlon Zambon,\authormark{2,*} A. Bieganowska,\authormark{3} F. Jabeen,\authormark{4,5} L. Vi\~na,\authormark{1,6,7} and C. Ant\'on-Solanas\authormark{1,6,7,*}}

\address{\authormark{1}Departamento de F\'isica de Materiales, Universidad Aut\'onoma de Madrid, 28049 Madrid, Spain\\
\authormark{2}Dipartimento di Fisica e Astronomia ``Galileo Galilei'', Universit\`a di Padova, I-35131 Padova, Italy\\
\authormark{3}Department of Experimental Physics, Faculty of Fundamental Problems of Technology, Wroc{\l}aw University of Science and Technology, Wyb.\ Wyspia{\'n}skiego 27, 50-370 Wroc{\l}aw, Poland.\\
\authormark{4}Institute of Physics, School of Basic Sciences, EPFL, 1015 Lausanne, Switzerland\\
\authormark{5}Present address: Univ Rennes, INSA Rennes, CNRS, Institut FOTON -- UMR 6082, F-35000 Rennes, France\\
\authormark{6}Instituto Nicol\'as Cabrera, Universidad Aut\'onoma de Madrid, 28049 Madrid, Spain\\
\authormark{7}Instituto de F\'isica de la Materia Condensada (IFIMAC), Universidad Aut\'onoma de Madrid, 28049 Madrid, Spain}

\email{\authormark{*}andreaherrero22@gmail.com,titta.carlonzambon@gmail.com,carlos.anton@uam.es}

\begin{abstract*}
We characterize spin-polarized lasing in a two-dimensional photonic lattice fabricated from a GaAs/InGaAs semiconductor microcavity sample. The lattice is defined by a staggered arrangement of rounded rectangular micrometric mesas that laterally confine and couple the optical modes. Polarization-, angle-, and energy-resolved micro-photoluminescence measurements reveal the transition from the strong-coupling regime to photon lasing, accompanied by extended spatial coherence across several lattice unit cells. Under circularly polarized nonresonant excitation, the emitted light acquires a controllable circular polarization whose handedness follows that of the pump. These results establish photonic-lattice VCSELs as a platform for spin-controlled coherent emission in extended optical systems.
\end{abstract*}

\section{\label{sec:intro} Introduction}

%The polarization of light provides a versatile degree of freedom for encoding, transmitting, and processing information in spin-based optical technologies. Spin lasers--devices where the polarization of emission is governed by the most populated spin of carriers in the gain medium--have emerged as promising platforms for information processing~\cite{Gerhardt2012_SpinVCSELs,pan_harnessing_2024}. In these systems, the optical selection rules of direct-gap semiconductors map the spin imbalance of electrons and holes directly onto the helicity of the emitted photons, enabling circularly polarized lasing. This spin-selective gain mechanism can lead to low lasing thresholds and ultrafast dynamics~\cite{Hsu2015_UltrafastSpinLaser,Lindemann2019,Labinac2025_BirefringentSpinLaser}, with demonstrations ranging from early spin-controlled VCSELs~\cite{Ando_photon-spin_1998,Gerhardt2012_SpinVCSELs} to self-polarized nanolasers~\cite{Chen2014_SelfPolarized} and complex devices incorporating anisotropies and exceptional points~\cite{Drong_spin-vcsels_2021,Drong2023_SpinVCSEL_EP,Almabetov_room_2024}. 

The polarization of light provides a versatile degree of freedom for encoding, transmitting, and processing information in spin-based optical technologies. Spin-lasers harness the ability to bias the spin-polarization of the gain medium to control the polarization of the emitted light of the carriers in the gain medium. In vertical cavity surface emitting lasers (VCSELs) based on direct-gap semiconductors, the optical selection rules map the spin imbalance of the carriers onto the helicity of the emitted photons \cite{Ando_photon-spin_1998,Gerhardt2012_SpinVCSELs,Lindemann2019}. Remarkably, the spin imbalance in the gain medium is nonlinearly amplified in the lasing regime, and can be controlled on picosecond timescales \cite{Damen1991,Munoz1995,Via1999,Gerlovin2004}, allowing ultrafast switching of the emission handedness \cite{martin_polarization_2002,Hsu2015_UltrafastSpinLaser,Lindemann2019}, with promising applications in the development of terahertz-bandwidth optical interconnect transceivers \cite{Lindemann2019,Drong_spin-vcsels_2021,Labinac2025_BirefringentSpinLaser}. 

From a more fundamental perspective, these devices exhibit a rich phase diagram, featuring exceptional points and self-oscillations \cite{Drong2023_SpinVCSEL_EP}. Moreover, when operated at cryogenic temperatures, exciton formation in high-finesse microcavity devices drives the system into the strong light--matter coupling regime \cite{Weisbuch1992}, giving rise to polariton quasi-particles \cite{Carusotto2013}. Microcavities in this regime have demonstrated optical spin injection, bistability, and spin jets~\cite{li_incoherent_2015,Pickup_polariton_2021,mirek_spin_2023}, as well as chiral emission induced by an optical Zeeman effect~\cite{Real2021_ChiralEmission}.
\begin{figure} 
\centering
\includegraphics[width=0.8\linewidth]{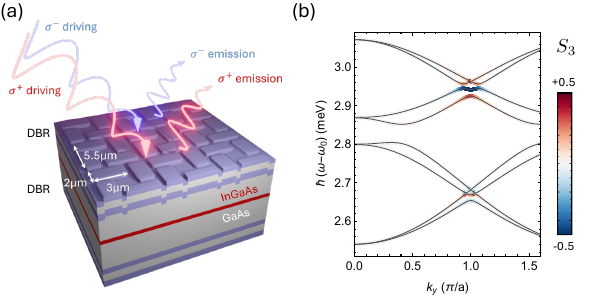}
\caption{\textbf{Sample design and simulated band structure of the staggered square lattice.} (a) Schematic of the semiconductor microcavity used in this work. The structure consists of a $\lambda$-thick GaAs cavity formed by two GaAs/AlAs distributed Bragg reflectors (20 and 24 DBR pairs) embedding a single In${0.06}$Ga${0.94}$As QW. The cavity spacer is patterned into a staggered square lattice of micrometer-scale mesas ($\sim 2 \times 3~\mu\mathrm{m}^2$) connected by narrow links that enable evanescent coupling between sites, forming a two-dimensional photonic lattice. Spin polarization of the gain medium via a circularly polarized nonresonant optical pump ($\sigma^\pm$) enables lasing in an extended mode with a well-defined emission handedness. (b) Calculated band structure (black lines) and corresponding degree of circular polarization of the photonic ($s$-type) modes of the lattice for $k_x = 0$, $\omega_0$ denotes the cutoff frequency of the planar cavity. The calculation relies on an effective optical Schr\"{o}dinger equation for the transverse lattice modes derived within the paraxial approximation. At the Brillouin zone edge, TE--TM splitting results in elliptically polarized modes, as indicated by degree of circular polarization $S_3$ (color coded).}
\label{fig:sketch_intro}
\end{figure}

VCSELs provide a reduced footprint, scalability, and high power-conversion efficiency \cite{Liu2019,pan_harnessing_2024}. However, the development of high-power, large-area devices remains challenging. Increasing the device size leads to a large number of spectrally overlapping transverse modes, which in turn induces multi-mode behavior mediated by spatial- and spectral-hole burning \cite{Giudici1998,VALLE1995297,Michalzik2013}, modulational instabilities \cite{Tai1986,Bobrovska2014,Baboux2018}, and chaos \cite{Mercadier2025}. As a result, large-area devices exhibit reduced spatial coherence \cite{Bobrovska2014,Baboux2018} and ultimately behave as an incoherent collection of smaller emitters. 

Two main approaches have been proposed to address this issue, both relying on transverse-mode engineering in coupled arrays of microlasers. The first is to trigger lasing in a topologically protected interface state of a VCSEL lattice \cite{Dikopoltsev2021,Ozawa2019}. The second is to engineer the photonic band structure of coupled microlaser arrays so that lasing occurs in states with an effective negative mass, thereby suppressing modulational instabilities \cite{Baboux2018,Fontaine2022}.

In the present work, we implement the latter strategy. Patterned microcavities enable transverse-mode engineering by imposing periodic potentials that determine the photon dispersion relation \cite{Carusotto2013,PhysRevB.74.155311}. Moreover, the inherent birefringence of distributed Bragg reflectors endows the platform with an effective spin--orbit coupling for photons \cite{Kavokin2005,Sala2015}, which can be exploited to tailor the polarization properties of the photonic modes \cite{Sala2015,Whittaker2018,Klembt2018,Widmann2025}. When linearly polarized bands hybridize due to spin--orbit coupling, extended circularly (or elliptically) polarized modes may emerge. Any spin imbalance in the gain medium can thus be harnessed to selectively promote lasing in co-circularly polarized optical modes \cite{CarlonZambon2019}.

Here, we demonstrate spin-polarized lasing in a two-dimensional VCSEL lattice supporting polariton quasiparticles in the linear regime. The lattice is fabricated from a GaAs/InGaAs microcavity patterned into a staggered array of mesas \cite{PhysRevB.74.155311}. Under circularly polarized nonresonant pumping, the emission undergoes a transition from the strong-coupling regime to photon lasing while maintaining a well-defined elliptical polarization dictated by the pump helicity. Above the VCSEL threshold, we observe extended spatial coherence across the excitation spot, evidencing the cooperative buildup of a spin-selective lasing state mediated by the lattice modes. Our results bridge two paradigms--spin-polarized lasing and extended modes in photonic lattices--and establish a pathway toward scalable coherent light sources with polarization injection. Beyond applications in spin-optoelectronics \cite{Gerhardt2012_SpinVCSELs,Lindemann2019}, this platform provides a versatile platform for exploring spin-dependent nonlinear dynamics and topological phases in coupled VCSEL arrays \cite{Smirnova2020}.

%[T]: note - i believe the limited degree of circular polarization obtained in the lasing regime is due to either both or one of these effects: 1) the state where we lase is a couple of orthogonal, energy degenerate linearly polarized modes, in this case the gain spin ibalance is mapped directly onto the imbalance of the emission (this could be checked with simulations but i fear the potential felt by photons in these measurements is not the designed one - the square lattice guess). Yet, the degree of spin-polarization has a nonlinear growth above threshold (fig 4), meaning some nonlinear interplay is going on. Then maybe the reason for a reduced S3 is that we need to pump really hard, probably electronic phase-space saturation and interaction become really strong and the depolarization rate grows nonlinearly, so that the effective spin imbalance of the gain medium decreases as power increases.

The work is organized as follows: In Sec.~\ref{sec:setup}, we describe the sample, together with the optical setup used for polarization-, angle-, and energy-resolved measurements. Sec.~\ref{sec:results} presents our experimental findings: we first analyze the evolution of the energy-momentum dispersion and the transition from strong-coupling to VCSEL regime with increasing pump power, then demonstrate the buildup of extended spatial coherence in the latter regime, and finally discuss the role of spin injection and polarization control in the photonic lattice. In Sec.~\ref{sec:conclusions}, we summarize the implications of our results for spin-polarized lasing in extended systems and their potential for spin-optoelectronic applications. 

\section{\label{sec:setup} Sample and experimental setup}

The sample consists of a semiconductor microcavity grown by molecular beam epitaxy on a GaAs substrate. The cavity is formed by two distributed Bragg reflectors (DBRs) composed of alternating $\lambda/4$ layers of GaAs/AlAs. The structure contains 20/24 top/bottom DBR pairs, respectively, enclosing a $\lambda$-thick GaAs cavity spacer. The spacer embeds a single In$_{0.06}$Ga$_{0.94}$As quantum well (QW) located at the antinode of the optical field, yielding a Rabi splitting of 3.35 meV (see Supplementary Material).

To engineer the transverse photonic modes of the system, the planar microcavity is patterned into micrometer-scale mesas using electron beam lithography followed by selective wet etching. Electron beam lithography was performed using hydrogen Silsesquioxqne (HSQ) as a negative tone resist. During the exposure and development, the HSQ pattern serves as a mask for wet chemical etching of GaAs using an NH$_4$OH:H$_2$O$_2$ solution. The etching process selectively removed GaAs while stopping at the embedded AlAs etch-stop layer. This process locally modifies the cavity thickness by about $6~\mathrm{nm}$, thereby lowering the resonance frequency inside the mesas and creating an effective potential well for photons \cite{ElDaif_2006,OuelletPlamondon2017}.

Arranging these mesas into periodic patterns enables the realization of photonic lattices in which confined modes of adjacent mesas couple evanescently. Here, we focus on a staggered square lattice formed by rounded rectangular mesas, see Fig.~\ref{fig:sketch_intro}(a). The mesas have lateral dimensions of approximately $2\times3\,\mu$m$^2$ and are separated by $0.5\,\mu$m. The resulting structure effectively defines a two-dimensional lattice potential for exciton-polaritons.

Figure~\ref{fig:sketch_intro}(b) shows the calculated photonic band structure of the staggered lattice. The calculation is based on an effective optical Schr\"{o}dinger equation describing the transverse modes of the lattice within the paraxial wave approximation (see supplementary material) \cite{Marte1997}. Since there are four mesas in the unit cell, and each uncoupled mesa supports two non-degenerate, linearly polarized ($s$-type) modes, eight bands arise from their hybridization. Polarization effects are included following Ref.~\cite{Kavokin2005}, using the TE--TM band splitting value $\Delta_{so}=0.05~\mathrm{meV}/\mu m^2$. Additional details on the simulation are provided in the Supplementary Materials. Interestingly, at the edge of the Brillouin zone, the calculated band structure exhibits the formation of elliptically polarized mode doublets. These doublets are non-degenerate in energy. However, their splitting is sufficiently small that it cannot be resolved within the linewidth or within the gain spectral envelope. The finite degree of circular polarization of the modes at the band edge ($S_3 \approx \pm 0.45 $) is essential for the implementation of a spin-laser via optical injection of spin-polarized carriers.

The optical characterization of the sample is performed using a micro--photoluminescence (PL) setup in a confocal configuration. The microcavity is non-resonantly excited with a continuous-wave Ti:Sapphire laser tuned to the first reflectivity minimum above the stop-band ($\sim 1.57$ eV). The excitation polarization is controlled using a linear polarizer and a quarter-wave plate (QWP). The emitted PL is collected through the same microscope objective and analyzed using real- and momentum-space imaging onto a spectrometer--CCD system, enabling energy- and angle-resolved measurements of the polariton dispersion. Polarization-resolved measurements use a rotating QWP and a linear polariser, providing the full Stokes parameters $S_i$, $i=1,2,3$ \cite{Wang2006_QWP,Wilkinson2021}. The characterization of the spatial-coherence is performed using a Michelson interferometer in a mirror retro-reflector configuration at zero delay. Further details and a sketch of the full experimental setup are provided in the Supplementary Material.
 
\section{\label{sec:results} Results}

\begin{figure}[htbp]
\centering
\includegraphics[width=0.95\linewidth]{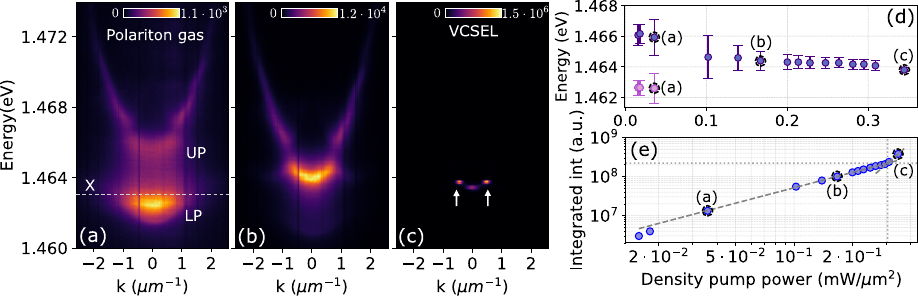}
\caption{ \textbf{Nonlinear input-output response.} (a--c) Energy- and momentum-resolved emission of the system as a function of increasing excitation power. 
(a) Low-density regime, showing the lower (LP) and upper (UP) polariton branches. The dashed white line in panel (a) indicates the exciton energy at $1.465 \pm 0.001~\text{eV}$. (b) Transition to the weak-coupling regime and onset of stimulated emission. (c) VCSEL regime. The arrows indicate the lasing from points in the dispersion relation with negative curvature. (d) Energy at the negative curvature points in the dispersion as a function of input power density; the error bars represent the full width at half maximum (FWHM) of the spectrum at these points. (e) Total integrated intensity measured as a function of pump power, showing a lasing threshold at $P_\text{th} \sim 0.3~\text{mW}/\mu\text{m}^2$ (vertical dashed gray line). All experiments were conducted using a large excitation spot ($\sim 34~\mu\text{m}$ FWHM), exciting about $5\times 5$ unit cells of the lattice.}
\label{fig:results_1}
\end{figure}

In this section, we present the experimental characterization of the staggered square polaritonic lattice under non-resonant optical excitation. Upon increasing the excitation power, we observe the onset of lasing. We analyze the corresponding changes in the energy--momentum dispersion, the development of spatial coherence across the lattice (under large-spot excitation), and the role of non-resonant spin injection in the emission polarization. Note that as the microcavity embeds a single QW, lasing occurs in the weak-coupling regime, hence polariton physics is not involved above threshold, and the sample can be regarded as a VCSEL.

\subsection{\label{subsec:pumppow} Energy-momentum dispersion relation and power dependence}

We investigate the PL of the staggered square lattice as a function of excitation power. Figure~\ref{fig:results_1}(a--c) shows the energy- and momentum-resolved PL of the system for three representative excitation powers. Figure~\ref{fig:results_1}(d) and (e) show, respectively, the emission energy and linewidth, and the integrated intensity characteristics of the sample as functions of excitation pump power density.

Initially, at low power density, the system is in the strong-coupling regime, as evidenced by the observation of both the upper and lower polariton branches in the energy--momentum dispersion (see Fig.~\ref{fig:results_1}(a)). The available sample presents mostly positive detunings, we focus our studies in a quadrant with a bare exciton--photon detuning of approximately zero. The lower polariton (LP) branch appears around $1.463~\text{eV}$ and features pronounced emission near $k = 0$.

Then, for excitation power densities exceeding $0.05~\text{mW}/\mu\text{m}^2$, we observe a gradual transition from the strong- to the weak-coupling regime, as indicated by the progressive reduction of the energy gap between the UP and LP branches (see Fig.~\ref{fig:results_1}(b)) and by linewidth broadening (see Fig.~\ref{fig:results_1}(d)). This behavior can be understood as follows: at increasing carrier densities, exciton--reservoir interactions broaden the emission linewidth, while phase-space filling effects progressively reduce the exciton oscillator strength, ultimately driving the system into the weak-coupling regime \cite{Kappei_2005,Amo2007-dk}. 

In the Supplementary Material, we present additional images of the dispersion relation across this transition. We also note that the progressive redshift of the emission peak observed in Fig.~\ref{fig:results_1}(d) above $0.2~\mathrm{mW}$ can be attributed to heating effects: both the QW bandgap and the effective refractive index of the microcavity decrease with increasing temperature \cite{Varshni_1967,Via1984}, likely due to the strong continuous-wave excitation.

Finally, above the threshold power density $P_{\text{th}} {\sim} 0.3~\text{mW}/\mu\text{m}^2$, we observe a sharp increase in emitted intensity, abrupt linewidth narrowing, and a redistribution of emission in momentum space (see Fig.~\ref{fig:results_1}(c--e)), evidencing the onset of lasing. In this regime, the sample can be regarded as a standard VCSEL. As shown in Fig.~\ref{fig:results_1}(c), lasing is not restricted to a single mode; multiple photonic modes become macroscopically occupied, resulting in multimode operation. 

Nonetheless, the strongest emission appears near points of negative curvature (see arrows in Fig.~\ref{fig:results_1}(c)), where extended modes can form free from modulational instabilities \cite{Baboux2018}, providing a key ingredient for achieving lasing in spin-polarized extended modes with optically defined handedness. Furthermore, single-mode operation may be achieved by tuning the exciton detuning relative to the bare cavity modes, optimizing the excitation profile, and implementing lattices with stronger lateral confinement \cite{Dikopoltsev2021,Fontaine2022}, for example using lithographically defined micropillar arrays.

\subsection{Extended coherence in the VCSEL regime}

In order to substantiate the claim that lasing in extended modes with large spatial coherence can be achieved for negative effective mass modes, we investigate the development of spatial coherence in our system via the first-order correlation function, $g^{(1)}(\mathbf{r},-\mathbf{r},\tau=0)$, using a Michelson interferometer in the retro-reflector configuration, allowing us to interfere emission from symmetric points $\mathbf{r}$ and $-\mathbf{r}$ at zero delay ($\tau=0$).
\begin{figure}[htbp]
\centering
\includegraphics[width=0.95\linewidth]{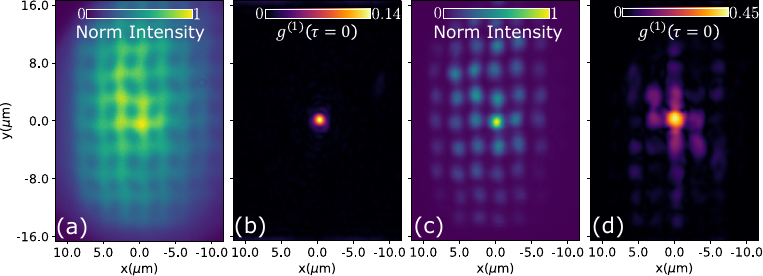}
\caption{\textbf{Spatial coherence via Michelson interferometry in an extended lattice.} Real space emission and $g^{(1)}$ measured in the polariton gas regime (a,b) and in the strong-coupling regime with a power of $0.29 P_{\textrm{th}}$ and (c,d) VCSEL regime with a power of $1.15 P_{\textrm{th}}$. The intensity distribution in panels (a,c) is normalized to unity. The real space map of $g^{(1)}$ is extracted via off-axis holographic techniques.}
\label{fig:interference_g1}
\end{figure}

Figure~\ref{fig:interference_g1} shows the emission intensity and first-order spatial coherence $g^{(1)}$ maps at two excitation regimes: low density and above the lasing threshold, corresponding to pump powers of $0.29 P_{\textrm{th}}$ and $1.15 P_{\textrm{th}}$ for the panels (a,b) and (c,d), respectively. At low excitation power, the emission is distributed evenly across all mesas within the pump spot (see Fig.~\ref{fig:interference_g1}(a)) and is spectrally broad, as shown in Fig.~\ref{fig:results_1}(a). In this regime, polaritons remain largely uncorrelated and incoherent, as expected when the population occupies many energy states with random phases. Consequently, the measured spatial coherence is essentially limited to a single lattice site.

Upon increasing the excitation power above the VCSEL threshold, the emission intensity in Fig.~\ref{fig:interference_g1} becomes slightly non-uniform due to multimode laser operation. However, a significant increase in the first-order coherence, reaching $g^{(1)}(\mathbf{r}{=}0) {\sim} 0.45$, is observed. This value is likely limited by the multi-mode character of the emission and by the near-threshold operation of the sample, even at the highest available excitation power. Nevertheless, the $g^{(1)}$ map clearly shows that coherence extends across multiple unit cells, providing strong evidence of lasing in an extended photonic mode.

Having characterized the input-output behavior of the sample, and that lasing in negative-curvature modes enables the buildup of extended spatial coherence, we now turn to characterizing the polarization properties of the PL emission.

%It is important to note that at high pump power the system enters the weak coupling regime, where photon lasing occurs instead of polariton condensation. In this regime, coherence arises purely from stimulated emission into a photonic mode and is not driven by polariton--polariton interactions. As such, we still observe the emergence of extended spatial coherence under VCSEL lasing. However, unlike polariton condensates, this coherence is not tied to the excitonic fraction or interaction-driven phase transitions, but rather to gain exceeding loss in the cavity mode. \red{[Titta, please, maybe you can close this section with some final statement. ]}

%[T]: If you want i can write something on the distinction between the emergence of coherence for a polariton laser and for a VCSEL but i feel we don't win so much in adding this discussion (which is touroughly discussed in the literature and not so necessary for the message of the paper) while actually triggering some question/clarification from referees 

\subsection{Spin injection in a photonic lattice}

The possibility of engineering extended optical modes with a well-defined circular polarization has been established mainly in chiral photonic architectures. For instance, in sub-wavelength GaAs waveguide structures patterned with chiral photonic-crystal slabs, the broken mirror symmetry of the unit cell leads to chiral modes characterized by unequal coupling to $\sigma^{\pm}$ radiation and, therefore, to finite circular polarization even in the absence of a magnetic field \cite{Maksimov2014,Lobanov1,Lobanov2}. The concept was later extended to room-temperature QW lasing in chiral microcavities \cite{Demenev2016}, and more recently to nano-imprinted perovskite-based devices operating under ambient conditions \cite{MendozaCarreo2023,FiuzaManeiro2024}. In these systems, the circular polarization is a direct consequence of the optically active chiral meta-unit cell. A key drawback of these devices is that the emission handedness is ultimately defined during fabrication, which is undesirable for spin opto-electronic applications \cite{Lindemann2019,Labinac2025_BirefringentSpinLaser,Drong_spin-vcsels_2021}.

By contrast, the micron-sized mesas in our staggered square lattice are not sub-wavelength chiral elements, and the inter-site coupling does not provide the optical polarization filtering characteristic of chiral structures with broken in-plane mirror symmetry. Consequently, although TE--TM splitting may still define an effective spin--orbit coupling, the geometry does not intrinsically favor one circular polarization over the other. Consistently, under linearly polarized excitation we observe no spontaneous handedness selection, and $S_3$ remains close to zero, indicating that the lattice does not imprint an intrinsic circular polarization on the emitted field (see Supplementary Material).

In principle, due to the $90^{\circ}$ rotations between neighboring mesas, a tight-binding description predicts no polarization projection between adjacent sites and, consequently, no geometry-induced spin--orbit coupling capable of lifting the degeneracy of the optical mode doublets \cite{Sala2015}. Nevertheless, as shown in Fig.~\ref{fig:sketch_intro}(b), band mixing occurs at the Brillouin zone edge, resulting in the formation of elliptically polarized mode doublets that are slightly split in energy. This behavior arises from the reduced confinement experienced by the modes of each mesa (compared to lithographically defined pillars) which enhances evanescent coupling to next-nearest neighbors, combined with the staggered arrangement of the mesas within the unit cell. Although the energy separation of these doublets is too small to be resolved within the emission linewidth or to selectively trigger lasing in a specific mode via spectral shaping of the gain profile, their elliptical polarization can, in principle, be exploited to preferentially promote lasing in the mode with the largest overlap with a partially spin-polarized gain medium \cite{CarlonZambon2019}. 

\begin{figure} 
\centering
\includegraphics[width=1\linewidth]{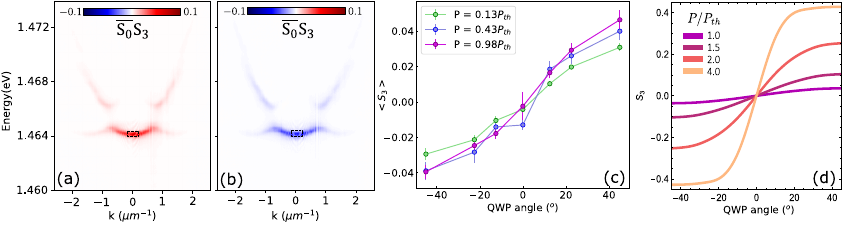}
\caption{\textbf{Spin injection in a photonic staggered-square lattice.} (a/b) Map of $\overline{S_0} S_3$ in the dispersion relation under $\sigma^\pm$ excitation at a fixed pump power density of $P=0.98 P_{\textrm{th}}$. The $S_0$ intensity map has been normalized to unity ($\overline{S_0}$) prior to the representation of the $\overline{S_0} S_3$ map, hence re-scaling its false color scale. The black rectangles in both panels indicate the region of interest where $S_3$ is averaged for the data points represented in the next panel. (c) Average $S_3$ as a function of the QWP angle of the excitation and the laser power density for three values, $0.13P_{\textrm{th}}$, $0.43P_{\textrm{th}}$ and $0.98P_{\textrm{th}}$. (d) Calculated degree of circular polarization of the emission as a function of the QWP angle according to the model in \eqref{eq:rate_SpinLaser} for increasing pump powers. The purple line corresponds to the data shown in panel (c) and yields a comparable value of $S_3$, with $S_3(45^\circ)=0.036$. Far above threshold, $S_3(45^\circ)$ saturates at the value $|\epsilon|\approx 0.45$. All model parameters are provided in the main text.}
\label{fig:spin_results}
\end{figure}

To gain further insight on the emission properties of the device, we examine the emission polarization as a function of both the pump power and the ellipticity of the non-resonant excitation. In particular, we focus on the degree of circular polarization ($S_3$) of the emission. Figure~\ref{fig:spin_results}(a,b) shows the product $\overline{S_0} S_3$ of the dispersion relation under $\sigma^\pm$ excitation at a pump power of $P =0.98P_{\textrm{th}}$. We could not explore higher pump powers in the input-output curve since we reached the maximum of the laser power accessible in our setup. The $S_0$ intensity map is normalized to unity ($\overline{S_0}$) prior to constructing the $\overline{S_0} S_3$ map. This representation highlights the degree of circular polarization only in regions of the dispersion with finite emission intensity. For the sake of completeness, a full Stokes vector reconstruction of the dispersion relation is provided in the Supplementary Material. 

Next, we study the $S_3$ value calculating its average $\langle S_3\rangle$ and standard deviation in a region at the bottom of the dispersion relation as a function of both the ellipticity and the power density of the excitation laser (see Fig.~\ref{fig:spin_results}(c)). Two main observations can be made. First, the strength of the $S_3$ component correlates with the excitation QWP angle, vanishing for horizontal linear polarization ($0^\circ$) and increasing progressively as the polarization becomes more elliptical, reaching a maximum near circular polarization injection ($\pm 45^{\circ}$), which is consistent with the fact that the spin-imbalance in the gain medium is responsible for the $S_3$ emission. Second, $S_3$ weakly grows with increasing excitation power for a fixed degree of pump polarization. This behavior originates from the spin imbalance created in the exciton reservoir by circularly polarized nonresonant excitation. While the degree of spin polarization of the gain medium is typically moderate ($10$--$15\%$) \cite{Ando_photon-spin_1998,Hsu2015_UltrafastSpinLaser,CarlonZambon2019} and tends to decrease at high optical power due to enhanced depolarization mechanisms \cite{Amand_1997}, near threshold, stimulated emission selectively amplifies the emission in the dominant spin channel through nonlinear mode competition, leading to an enhanced $S_3$ of the output light \cite{Ostermann2013}. We note that the points of $\langle S_3\rangle$ in Fig. \ref{fig:spin_results}(c) present a slightly smaller value than the overall value of the $S_3$ maps shown in the supplement, as a result of the integration in a given area of the dispersion relation.

A minimal model of the nonlinear dynamics of the system can be derived by assuming that lasing occurs only in a doublet of optical modes at the Brillouin zone edge, characterized by finite degrees of circular polarization $\pm \epsilon$ and intensities $I_{\pm}$, see Fig.~\ref{fig:sketch_intro}(b). Optical pumping can be used to inject spin-polarized carriers, whose populations are described by $N_{\sigma_{\pm}}$ and are subject to depolarization during the relaxation process (at a rate $\gamma_s$). In the Supplementary Material, we show how to derive the steady-state rate equations for this minimal two-mode spin-laser model in terms of the total intensity $S_0=I_{+}+I_{-}$, the degree of elliptical polarization $S_{\epsilon}=(I_{+}-I_{-})/S_0$ ($S_3=\epsilon S_{\epsilon}$), the total carrier population $N_0=N_{\sigma_{+}}+N_{\sigma_{-}}$, and the degree of spin imbalance $\Pi_{s}=(N_{\sigma_{+}}-N_{\sigma_{-}})/N_0$. The steady-state equations read:
\begin{equation}\label{eq:rate_SpinLaser}
\begin{aligned}
    0 &= -\kappa S_0 + \beta\Gamma N_0 + \frac{g_0 S_0 N_0}{2}(1+\epsilon S_{\epsilon}\Pi_s)\\
    0 &= P_0 -\Gamma N_0 - \frac{g_0 S_0 N_0}{2}(1+\epsilon S_{\epsilon}\Pi_s)\\
    0 &= -\beta\Gamma N_0 (S_\epsilon - \epsilon \Pi_s) +\frac{g_0 N_0 S_0}{2} \epsilon\Pi_s (1-S_\epsilon^2)\\
    0 & = P_0 (\alpha-\Pi_s) - 2 \gamma_s N_0\Pi_s  -\frac{g_0 N_0 S_0}{2}\epsilon S_{\epsilon} (1-\Pi_s^2).
\end{aligned}
\end{equation}
where $\kappa^{-1} \approx 10 ~\text{ps}$ is the cavity decay time, $\Gamma/\kappa=0.1$ and $\gamma_s/\Gamma = 4$ denote the carrier and spin relaxation rates, respectively, $\alpha=2/3$ is determined by the optical selection rules of GaAs QWs, $g_0 / \kappa = 0.01$ is the unsaturated gain coefficient, and $\beta=0.01$ is the laser beta factor. Figure \ref{fig:spin_results}(d) shows the calculated dependence of $S_3$ on the degree of circular polarization of the optical pump. We observe that the model captures the main features observed experimentally. Interestingly, the model predicts that, far above threshold, the degree of circular polarization of the PL emission is limited only by the underlying $S_3=\epsilon$ value of the optical modes. In other words, the model shows that, by switching the circular polarization of the optical pump, it is possible to selectively trigger lasing in either $I_{+}$ or $I_{-}$, similarly to Refs.~\cite{Ando_photon-spin_1998,Hsu2015_UltrafastSpinLaser,Lindemann2019,CarlonZambon2019}. Further details on the model can be found in the Supplementary Materials. Note that we did not include gain saturation effects, which may lead to an even richer phenomenology, such as bistability and chaos.

% \red{[TO DO] Once we have a final version of the figure, add the model and comment the two}

%The circular Stokes component $S_3$ should increase with excitation power as the system approaches the photon-lasing threshold. This behaviour originates from the spin imbalance created in the exciton reservoir by circularly polarized nonresonant excitation. The resulting spin-polarized carrier population provides spin-dependent gain for the cavity modes. Near threshold, stimulated emission selectively amplifies the dominant spin channel, leading to an enhanced degree of circular polarization in the emitted light \cite{CarlonZambon2019}.

%We observe that: (1) the spin injection increases significantly at higher powers, indicating enhanced spin preservation as the system approaches the photonic lasing threshold; and (2) the strength of the $S_3$ component correlates with the excitation quarter wave-plate angle, vanishing for horizontal linear polarization ($0^\circ$) and increasing progressively as the polarization becomes more elliptical, reaching a maximum near circular polarization ($\pm 45^{\circ}$). This behavior demonstrates that even a substantial structural perturbation to the microcavity, such as the introduction of a staggered square lattice, does not erase the spin-injection retained in the exciton reservoir.

\section{\label{sec:conclusions} Conclusions}

We have demonstrated spin-polarized lasing in a two-dimensional photonic lattice fabricated from a GaAs/InGaAs semiconductor microcavity. The staggered-square lattice provides transverse-mode engineering over an extended area, while preserving the spin-selective optical response of the embedded QW gain medium.

At low excitation power, angle-resolved PL reveals the lower and upper polariton branches of the patterned microcavity, confirming operation in the strong-coupling regime. As the excitation density is increased, the system undergoes a progressive transition to photon lasing above threshold. In this regime, the emission displays linewidth narrowing, a nonlinear increase in intensity, and occupation of several photonic lattice modes, with prominent emission from regions of negative curvature in the dispersion.

Interferometric measurements show that the lasing transition is accompanied by the buildup of spatial coherence extending over several lattice sites. This demonstrates that, despite the multimode character of the emission, the patterned VCSEL lattice supports coherent emission over an area larger than a single mesa, consistent with lasing in extended lattice modes.

Finally, polarization-resolved measurements show that circularly polarized nonresonant excitation produces a controllable circular component in the emitted light. The sign of the emission helicity follows that of the pump, whereas linearly polarized excitation does not lead to spontaneous handedness selection. This confirms that the observed circular polarization originates primarily from non-resonant spin injection into the gain medium.

Overall, these results establish patterned semiconductor microcavities as a versatile platform for combining transverse-mode engineering, extended spatial coherence, and spin-controlled photon lasing. They provide a route toward scalable spin-VCSEL arrays and, more broadly, toward photonic lattices in which gain, polarization, and spin--orbit coupling can be jointly engineered for nonlinear and topological photonic applications. Interestingly, our results also suggest the possibility of triggering lasing in topological edge states of lattices featuring SOC effects, with a propagation direction that can be optically controlled in the absence of external magnetic fields.

\begin{backmatter}

\bmsection{Funding}
Ministerio de Ciencia e Innovaci\'on (PID2023-148061NB-I00, PCI2024-153425, CNS2025-165107); Fundaci\'on Ram\'on Areces (ULTRABRIGHT); Mar\'ia de Maeztu Program for Units of Excellence in R\&D (CEX2023-001316-M); Fundaci\'on BBVA, Leonardo Grants for Researchers in Physics 2023; Comunidad de Madrid, Atracci\'on de Talento Mod. 1 (2020-T1/IND-19785).

\bmsection{Acknowledgments}
A.H.O. acknowledges the ``Beca de Colaboraci\'on 2024-2025" from the Ministerio de Educación y Formación Profesional and the fellowship from the ``La Caixa'' Foundation (ID 100010434), code ``PFA25-02207''. A.B. acknowledges support from the Erasmus+ programme through a recent graduate traineeship carried out at the Universidad Aut\'onoma de Madrid. L.V. and C.A.-S. acknowledge the support from the projects from the Ministerio de Ciencia e Innovaci\'on PID2023-148061NB-I00, PCI2024-153425 and CNS2025-165107, the project ULTRABRIGHT from the Fundaci\'on Ram\'on Areces and the ``Mar\'ia de Maeztu" Program for Units of Excellence in R\&D (CEX2023-001316-M). C. A.-S. acknowledges the Grant ``Leonardo for researchers in Physics 2023" from Fundaci\'on BBVA and the support from the Comunidad de Madrid fund ``Atracci\'on de Talento, Mod. 1", Ref. 2020-T1/IND-19785.

\bmsection{Disclosures}
The authors declare no conflicts of interest.

\bmsection{Data availability}
Data underlying the results presented in this paper will be made available in the e-cienciaDatos repository. The repository identifier should be inserted before submission.

\bmsection{Supplemental document}
See Supplement 1 for supporting content.

\end{backmatter}

\bibliography{references}

\begin{thebibliography}{99}
\bibitem{Gerhardt2012_SpinVCSELs} Nils C. Gerhardt, Martin R. Hofmann. ``Spin-Controlled Vertical-Cavity Surface-Emitting Lasers," \textit{Advances in Optical Technologies} \textbf{2012}, 268949 (2012). doi: 10.1155/2012/268949.
\bibitem{pan_harnessing_2024} Guanzhong Pan, Meng Xun, Xiaoli Zhou, Yun Sun, Yibo Dong, Dexin Wu. ``Harnessing the capabilities of VCSELs: unlocking the potential for advanced integrated photonic devices and systems," \textit{Light: Science \& Applications} \textbf{13}(1), 229 (2024). doi: 10.1038/s41377-024-01561-8.
\bibitem{Hsu2015_UltrafastSpinLaser} Feng-kuo Hsu, Wei Xie, Yi-Shan Lee, Sheng-Di Lin, Chih-Wei Lai. ``Ultrafast spin-polarized lasing in a highly photoexcited semiconductor microcavity at room temperature," \textit{Phys. Rev. B} \textbf{91}, 195312 (2015). doi: 10.1103/PhysRevB.91.195312.
\bibitem{Lindemann2019} Lindemann, Markus, Xu, Gaofeng, Pusch, Tobias, Michalzik, Rainer, Hofmann, Martin R., \v{Z}uti\'c, Igor, Gerhardt, Nils C.. ``Ultrafast spin-lasers," \textit{Nature} \textbf{568}(7751), 212--215 (2019). doi: 10.1038/s41586-019-1073-y.
\bibitem{Labinac2025_BirefringentSpinLaser} Velimir Labinac, Jiayu David Cao, Gaofeng Xu, Igor {\v Z}uti{\'c}. ``Birefringent spin-laser as a system of coupled harmonic oscillators," \textit{Phys. Rev. B} \textbf{112}, 075303 (2025). doi: 10.1103/jvwg-l6jw.
\bibitem{Ando_photon-spin_1998} H. Ando, T. Sogawa, H. Gotoh. ``Photon-spin controlled lasing oscillation in surface-emitting lasers," \textit{Appl. Phys. Lett.} \textbf{73}(5), 566--568 (1998). doi: 10.1063/1.121857.
\bibitem{Chen2014_SelfPolarized} Ju-Ying Chen, Tong-Ming Wong, Che-Wei Chang, Chen-Yuan Dong, Yang-Fang Chen. ``Self-polarized spin-nanolasers," \textit{Nat. Nanotechnol.} \textbf{9}, 845--850 (2014). doi: 10.1038/nnano.2014.195.
\bibitem{Drong_spin-vcsels_2021} M. Drong, T. F{\"o}rd{\"o}s, H. Y. Jaffr{\`e}s, J. Pe{\v r}ina, K. Postava, P. Ciompa, J. Pi{\v s}tora, H.-J. Drouhin. ``Spin-VCSELs with Local Optical Anisotropies: Toward Terahertz Polarization Modulation," \textit{Phys. Rev. Applied} \textbf{15}(1), 014041 (2021). doi: 10.1103/PhysRevApplied.15.014041.
\bibitem{Drong2023_SpinVCSEL_EP} Mariusz Drong, Jan Pe{\v r}ina Jr., Tibor F{\"o}rd{\"o}s, Henri Y. Jaffr{\`e}s, Kamil Postava, Henri-Jean Drouhin. ``Spin vertical-cavity surface-emitting lasers with linear gain anisotropy: Prediction of exceptional points and nontrivial dynamical regimes," \textit{Phys. Rev. A} \textbf{107}, 033509 (2023). doi: 10.1103/PhysRevA.107.033509.
\bibitem{Almabetov_room_2024} Almabetov, Timur, Androvitsaneas, Petros, Ren, Zhongze, Young, Andrew, Oulton, Ruth, Harbord, Edmund. ``Room temperature spin injection into commercial VCSELs at non-resonant wavelengths," \textit{arXiv} (2025). doi: 10.48550/ARXIV.2506.18376.
\bibitem{Damen1991} Damen, T. C., Vi\~na, Luis, Cunningham, J. E., Shah, Jagdeep, Sham, L. J.. ``Subpicosecond spin relaxation dynamics of excitons and free carriers in GaAs quantum wells," \textit{Physical Review Letters} \textbf{67}(24), 3432--3435 (1991). doi: 10.1103/physrevlett.67.3432.
\bibitem{Munoz1995} Mu\~noz, L., P\'erez, E., Vi\~na, L., Ploog, K.. ``Spin relaxation in intrinsic GaAs quantum wells: Influence of excitonic localization," \textit{Phys. Rev. B} \textbf{51}, 4247--4257 (1995). doi: 10.1103/PhysRevB.51.4247.
\bibitem{Via1999} Vi\~na, L. ``Spin relaxation in low-dimensional systems," \textit{Journal of Physics: Condensed Matter} \textbf{11}(31), 5929--5952 (1999). doi: 10.1088/0953-8984/11/31/304.
\bibitem{Gerlovin2004} Gerlovin, I. Ya., Dolgikh, Yu. K., Eliseev, S. A., Ovsyankin, V. V., Efimov, Yu. P., Ignatiev, I. V., et al.. ``Spin dynamics of carriers in GaAs quantum wells in an external electric field," \textit{Phys. Rev. B} \textbf{69}, 035329 (2004). doi: 10.1103/PhysRevB.69.035329.
\bibitem{martin_polarization_2002} M. D. Mart{\'\i}n, G. Aichmayr, L. Vi{\~n}a, R. Andr{\'e}. ``Polarization Control of the Nonlinear Emission of Semiconductor Microcavities," \textit{Phys. Rev. Lett.} \textbf{89}(7), 077402 (2002). doi: 10.1103/PhysRevLett.89.077402.
\bibitem{Weisbuch1992} C. Weisbuch, M. Nishioka, A. Ishikawa, Y. Arakawa. ``Observation of the coupled exciton-photon mode splitting in a semiconductor quantum microcavity," \textit{Phys. Rev. Lett.} \textbf{69}, 3314 (1992).
\bibitem{Carusotto2013} Carusotto, Iacopo, Ciuti, Cristiano. ``Quantum fluids of light," \textit{Rev. Mod. Phys.} \textbf{85}, 299--366 (2013). doi: 10.1103/RevModPhys.85.299.
\bibitem{li_incoherent_2015} G. Li, T. C. H. Liew, O. A. Egorov, E. A. Ostrovskaya. ``Incoherent excitation and switching of spin states in exciton-polariton condensates," \textit{Phys. Rev. B} \textbf{92}(6), 064304 (2015). doi: 10.1103/PhysRevB.92.064304.
\bibitem{Pickup_polariton_2021} L. Pickup, J. D. T{\"o}pfer, H. Sigurdsson, P. G. Lagoudakis. ``Polariton spin jets through optical control," \textit{Phys. Rev. B} \textbf{103}(15), 155302 (2021). doi: 10.1103/PhysRevB.103.155302.
\bibitem{mirek_spin_2023} R. Mirek, M. Furman, M. Kr{\'o}l, B. Seredy{\'n}ski, K. {\L}empicka-Mirek, K. Tyszka, et al.. ``Spin polarization of exciton-polariton condensate in a photonic synthetic effective magnetic field," \textit{Phys. Rev. B} \textbf{107}(12), 125303 (2023). doi: 10.1103/PhysRevB.107.125303.
\bibitem{Real2021_ChiralEmission} B. Real, N. Carlon Zambon, P. St-Jean, I. Sagnes, A. Lema{\^\i}tre, A. Harouri, et al.. ``Chiral emission induced by optical Zeeman effect in polariton micropillars," \textit{Phys. Rev. Research} \textbf{3}, 043161 (2021). doi: 10.1103/PhysRevResearch.3.043161.
\bibitem{Liu2019} Anjin Liu, Philip Wolf, James A. Lott, Dieter Bimberg. ``Vertical-cavity surface-emitting lasers for data communication and sensing," \textit{Photon. Res.} \textbf{7}(2), 121--136 (2019). doi: 10.1364/PRJ.7.000121.
\bibitem{Giudici1998} Giudici, M., Tredicce, J.R., Vaschenko, G., Rocca, J.J., Menoni, C.S.. ``Spatio-temporal dynamics in vertical cavity surface emitting lasers excited by fast electrical pulses," \textit{Optics Communications} \textbf{158}(1-6), 313--321 (1998). doi: 10.1016/S0030-4018(98)00326-5.
\bibitem{VALLE1995297} Valle, A, Sarma, J, Shore, K.A.. ``Dynamics of transverse mode competition in vertical cavity surface emitting laser diodes," \textit{Optics Communications} \textbf{115}(3-4), 297--302 (1995). doi: 10.1016/0030-4018(94)00707-2.
\bibitem{Michalzik2013} Michalzik, Rainer. ``VCSELs," \textit{Springer Berlin Heidelberg} \textbf{166}, 560 (2013). doi: 10.1007/978-3-642-24986-0.
\bibitem{Tai1986} Tai, K., Hasegawa, A., Tomita, A.. ``Observation of modulational instability in optical fibers," \textit{Physical Review Letters} \textbf{56}(2), 135--138 (1986). doi: 10.1103/PhysRevLett.56.135.
\bibitem{Bobrovska2014} Bobrovska, Nataliya, Ostrovskaya, Elena A., Matuszewski, Micha{\l}. ``Stability and spatial coherence of nonresonantly pumped exciton-polariton condensates," \textit{Physical Review B} \textbf{90}(20), 205304 (2014). doi: 10.1103/PhysRevB.90.205304.
\bibitem{Baboux2018} Baboux, F., {De Bernardis}, D., Goblot, V., Gladilin, V. N., Gomez, C., Galopin, E., et al.. ``Unstable and stable regimes of polariton condensation," \textit{Optica} \textbf{5}(10), 1163 (2018). doi: 10.1364/optica.5.001163.
\bibitem{Mercadier2025} Mercadier, Jules, Bittner, Stefan, Rontani, Damien, Sciamanna, Marc. ``Chaos from a free-running broad-area VCSEL," \textit{Optics Letters} \textbf{50}(3), 796 (2025). doi: 10.1364/OL.545404.
\bibitem{Dikopoltsev2021} Dikopoltsev, Alex, Harder, Tristan H, Lustig, Eran, Egorov, Oleg A, Beierlein, Johannes, Wolf, Adriana, et al.. ``Topological insulator vertical-cavity laser array," \textit{Science} \textbf{373}(6562), 1514--1517 (2021). doi: 10.1126/science.abj2232.
\bibitem{Ozawa2019} Ozawa, Tomoki, Price, Hannah M., Amo, Alberto, Goldman, Nathan, Hafezi, Mohammad, Lu, Ling, et al.. ``Topological photonics," \textit{Rev. Mod. Phys.} \textbf{91}, 015006 (2019). doi: 10.1103/RevModPhys.91.015006.
\bibitem{Fontaine2022} Fontaine, Quentin, Squizzato, Davide, Baboux, Florent, Amelio, Ivan, Lema{\^{i}}tre, Aristide, Morassi, Martina, et al.. ``Kardar--Parisi--Zhang universality in a one-dimensional polariton condensate," \textit{Nature} \textbf{608}(7924), 687--691 (2022). doi: 10.1038/s41586-022-05001-8.
\bibitem{PhysRevB.74.155311} R. Idrissi Kaitouni, O. El Da{\"i}f, A. Baas, M. Richard, T. Paraiso, P. Lugan, et al.. ``Engineering the spatial confinement of exciton polaritons in semiconductors," \textit{Phys. Rev. B} \textbf{74}(15), 155311 (2006). doi: 10.1103/PhysRevB.74.155311.
\bibitem{Kavokin2005} Kavokin, Alexey, Malpuech, Guillaume, Glazov, Mikhail. ``Optical Spin Hall Effect," \textit{Physical Review Letters} \textbf{95}(13), 136601 (2005). doi: 10.1103/PhysRevLett.95.136601.
\bibitem{Sala2015} V. G. Sala, D. D. Solnyshkov, I. Carusotto, T. Jacqmin, A. Lema{\^\i}tre, H. Ter{\c c}as, et al.. ``Spin-Orbit Coupling for Photons and Polaritons in Microstructures," \textit{Phys. Rev. X} \textbf{5}, 011034 (2015). doi: 10.1103/PhysRevX.5.011034.
\bibitem{Whittaker2018} Whittaker, C. E., Cancellieri, E., Walker, P. M., Gulevich, D. R., Schomerus, H., Vaitiekus, D., et al.. ``Exciton Polaritons in a Two-Dimensional Lieb Lattice with Spin-Orbit Coupling," \textit{Physical Review Letters} \textbf{120}(9), 097401 (2018). doi: 10.1103/PhysRevLett.120.097401.
\bibitem{Klembt2018} S. Klembt, T. H. Harder, O. A. Egorov, K. Winkler, R. Ge, M. A. Bandres, et al.. ``Exciton-polariton topological insulator," \textit{Nature} \textbf{562}(7728), 552--556 (2018). doi: 10.1038/s41586-018-0601-5.
\bibitem{Widmann2025} Simon Widmann, Jonas Bellmann, Johannes D\"ureth, Siddhartha Dam, Christian G. Mayer, Philipp Gagel, et al.. ``Artificial Gauge Fields and Dimensions in a Polariton Hofstadter Ladder,"  (2025). arXiv:2506.13521.
\bibitem{CarlonZambon2019} {Carlon Zambon}, N., St-Jean, P., Mili{\'{c}}evi{\'{c}}, M., Lema{\^{i}}tre, A., Harouri, A., {Le Gratiet}, L., et al.. ``Optically controlling the emission chirality of microlasers," \textit{Nature Photonics} \textbf{13}(4), 283--288 (2019). doi: 10.1038/s41566-019-0380-z.
\bibitem{Smirnova2020} Smirnova, Daria, Leykam, Daniel, Chong, Yidong, Kivshar, Yuri. ``Nonlinear topological photonics," \textit{Applied Physics Reviews} \textbf{7}(2), 021306 (2020). doi: 10.1063/1.5142397.
\bibitem{ElDaif_2006} El Da\"if, O., Baas, A., Guillet, T., Brantut, J.-P., Kaitouni, R. Idrissi, Staehli, J. L., Morier-Genoud, F., Deveaud, B.. ``Polariton quantum boxes in semiconductor microcavities," \textit{Applied Physics Letters} \textbf{88}(6), 061105 (2006). doi: 10.1063/1.2172409.
\bibitem{OuelletPlamondon2017} Claud{\'e}ric Ouellet-Plamondon. ``On the Physics of Multimode Polaritons,"  (2017). doi: 10.5075/EPFL-THESIS-7603.
\bibitem{Marte1997} Marte, Monika A. M., Stenholm, Stig. ``Paraxial light and atom optics: The optical Schr\"odinger equation and beyond," \textit{Phys. Rev. A} \textbf{56}, 2940--2953 (1997). doi: 10.1103/PhysRevA.56.2940.
\bibitem{Wang2006_QWP} Wang, Zheng Ping, Li, Qing Bo, Tan, Qiao, Huang, Zong Jun, Shi, Jin Hui. ``Method of measuring the practical retardance and judging the fast or slow axis of a quarter-wave plate," \textit{Measurement} \textbf{39}(8), 729--735 (2006). doi: 10.1016/j.measurement.2006.03.005.
\bibitem{Wilkinson2021} Wilkinson, T A, Maurer, C E, Flood, C J, Lander, G, Chafin, S, Flagg, E B. ``Complete Stokes vector analysis with a compact, portable rotating waveplate polarimeter," \textit{Rev. Sci. Instrum.} \textbf{92}(9), 093101 (2021).
\bibitem{Kappei_2005} Kappei, L., Szczytko, J., Morier-Genoud, F., Deveaud, B.. ``Direct Observation of the Mott Transition in an Optically Excited Semiconductor Quantum Well," \textit{Phys. Rev. Lett.} \textbf{94}, 147403 (2005). doi: 10.1103/PhysRevLett.94.147403.
\bibitem{Amo2007-dk} Amo, A, Mart{\'\i}n, M D, Vi{\~n}a, L, Toropov, A I, Zhuravlev, K S. ``Photoluminescence dynamics in {GaAs} along an optically induced Mott transition," \textit{J. Appl. Phys.} \textbf{101}(8), 081717 (2007).
\bibitem{Varshni_1967} Y.P. Varshni. ``Temperature dependence of the energy gap in semiconductors," \textit{Physica} \textbf{34}(1), 149-154 (1967). doi: https://doi.org/10.1016/0031-8914(67)90062-6.
\bibitem{Via1984} Vi\~na, L., Logothetidis, S., Cardona, M.. ``Temperature dependence of the dielectric function of germanium," \textit{Physical Review B} \textbf{30}(4), 1979--1991 (1984). doi: 10.1103/physrevb.30.1979.
\bibitem{Maksimov2014} Maksimov, A. A., Tartakovskii, I. I., Filatov, E. V., Lobanov, S. V., Gippius, N. A., Tikhodeev, S. G., et al.. ``Circularly polarized light emission from chiral spatially-structured planar semiconductor microcavities," \textit{Physical Review B} \textbf{89}(4) (2014). doi: 10.1103/physrevb.89.045316.
\bibitem{Lobanov1} Lobanov, S. V., Tikhodeev, S. G., Gippius, N. A., Maksimov, A. A., Filatov, E. V., Tartakovskii, I. I., et al.. ``Controlling circular polarization of light emitted by quantum dots using chiral photonic crystal slabs," \textit{Physical Review B} \textbf{92}(20) (2015). doi: 10.1103/physrevb.92.205309.
\bibitem{Lobanov2} Lobanov, Sergey V., Weiss, Thomas, Gippius, Nikolay A., Tikhodeev, Sergei G., Kulakovskii, Vladimir D., Konishi, Kuniaki, Kuwata-Gonokami, Makoto. ``Polarization control of quantum dot emission by chiral photonic crystal slabs," \textit{Optics Letters} \textbf{40}(7), 1528 (2015). doi: 10.1364/ol.40.001528.
\bibitem{Demenev2016} Demenev, A. A., Kulakovskii, V. D., Schneider, C., Brodbeck, S., Kamp, M., H\"{o}fling, S., et al.. ``Circularly polarized lasing in chiral modulated semiconductor microcavity with GaAs quantum wells," \textit{Applied Physics Letters} \textbf{109}(17) (2016). doi: 10.1063/1.4966279.
\bibitem{MendozaCarreo2023} Mendoza-Carre\~no, Jose, Molet, Pau, Otero-Mart\'inez, Clara, Alonso, Maria Isabel, Polavarapu, Lakshminarayana, Mihi, Agust\'in. ``Nanoimprinted 2D-Chiral Perovskite Nanocrystal Metasurfaces for Circularly Polarized Photoluminescence," \textit{Advanced Materials}, 2210477 (2023). doi: 10.1002/adma.202210477.
\bibitem{FiuzaManeiro2024} Fiuza-Maneiro, Nadesh, Mendoza-Carre\~no, Jose, G\'omez-Gra\~na, Sergio, Alonso, Maria Isabel, Polavarapu, Lakshminarayana, Mihi, Agust\'in. ``Inducing Efficient and Multiwavelength Circularly Polarized Emission From Perovskite Nanocrystals Using Chiral Metasurfaces," \textit{Advanced Materials} \textbf{36}(52) (2024). doi: 10.1002/adma.202413967.
\bibitem{Amand_1997} Amand, T., Robart, D., Marie, X., Brousseau, M., Le Jeune, P., Barrau, J.. ``Spin relaxation in polarized interacting exciton gas in quantum wells," \textit{Phys. Rev. B} \textbf{55}, 9880--9896 (1997). doi: 10.1103/PhysRevB.55.9880.
\bibitem{Ostermann2013} Johannes Michael Ostermann, Rainer Michalzik. ``Polarization Control of VCSELs," \textit{VCSELs: Fundamentals, Technology and Applications of Vertical-Cavity Surface-Emitting Lasers}, 147--179 (2013). doi: 10.1007/978-3-642-24986-0\_5.
\end{thebibliography}

\end{document}